\documentclass{article}

\usepackage{epsf,latexsym,amssymb}
\usepackage[mathscr]{eucal} 

\usepackage{amsmath}  

\newcommand{\qed}{\hfill \halmos} 

\newcommand{\R}{{\mathbb R}}  

\newtheorem{theorem}{Theorem}
\newtheorem{itlemma}{Lemma}[section] 
\newtheorem{itproposition}[itlemma]{Proposition}
\newtheorem{itcorollary}[itlemma]{Corollary}
\newtheorem{itremark}[itlemma]{Remark}
\newtheorem{itdefinition}[itlemma]{Definition}
\newtheorem{itexample}[itlemma]{Example}

\newenvironment{lemma}{\begin{itlemma}\rm}{\end{itlemma}} 
\newenvironment{remark}{\begin{itremark}\rm}{\end{itremark}} 
\newenvironment{corollary}{\begin{itcorollary}\rm}{\end{itcorollary}}
\newenvironment{proposition}{\begin{itproposition}\rm}{\end{itproposition}}
\newenvironment{definition}{\begin{itdefinition}\rm}{\end{itdefinition}}
\newenvironment{example}{\begin{itexample}\rm}{\end{itexample}}

\newcommand{\be}[1]{\begin{equation}\label{#1}}
\newcommand{\ee}{\end{equation}}
\newcommand{\bl}[1]{\begin{lemma}\label{#1}}
\newcommand{\ble}[1]{\begin{lemmaex}\label{#1}}
\newcommand{\br}[1]{\begin{remark}\label{#1}}
\newcommand{\bt}[1]{\begin{theorem}\label{#1}}
\newcommand{\bd}[1]{\begin{definition}\label{#1}}
\newcommand{\bp}[1]{\begin{proposition}\label{#1}}
\newcommand{\bc}[1]{\begin{corollary}\label{#1}}
\newcommand{\bfact}[1]{\begin{fact}\label{#1}}
\newcommand{\ber}[1]{\begin{exercise}\label{#1}}
\newcommand{\bex}[1]{\begin{example}\label{#1}}
\newcommand{\bem}[1]{\begin{example}\label{#1}}  
\newcommand{\ec}{\mybox\end{corollary}}
\newcommand{\efact}{\mybox\end{fact}}
\newcommand{\eer}{\mybox\end{exercise}}
\newcommand{\eex}{\mybox\end{example}}
\newcommand{\eem}{\mybox\end{example}}
\newcommand{\el}{\mybox\end{lemma}}
\newcommand{\ele}{\mybox\end{lemmaex}}
\newcommand{\er}{\mybox\end{remark}}
\newcommand{\et}{\qed\end{theorem}}
\newcommand{\ed}{\mybox\end{definition}}
\newcommand{\ep}{\mybox\end{proposition}}
\newcommand{\epr}{\end{proof}}
\newcommand{\bpr}{\begin{proof}}

\newcommand{\ecs}{\end{corollary}}
\newcommand{\eers}{\end{exercise}}
\newcommand{\eexs}{\end{example}}
\newcommand{\eems}{\end{example}}
\newcommand{\els}{\end{lemma}}
\newcommand{\eles}{\end{lemmaex}}
\newcommand{\ers}{\end{remark}}
\newcommand{\ets}{\end{theorem}}
\newcommand{\eds}{\end{definition}}
\newcommand{\eps}{\end{proposition}}
\newcommand{\halmos}{\rule{1ex}{1.4ex}}
\newcommand{\mybox}{\hfill $\Box$} 
\newcommand{\beq}{\begin{eqnarray}}
\newcommand{\eeq}{\end{eqnarray}}
\newcommand{\beqn}{\begin{eqnarray*}}
\newcommand{\eeqn}{\end{eqnarray*}}
\newcommand{\bi}{\begin{itemize}}
\newcommand{\ei}{\end{itemize}}
\newcommand{\ben}{\begin{enumerate}}
\newcommand{\een}{\end{enumerate}}

\newenvironment{proof}{\noindent {\em Proof}.\ }{\hspace*{\fill}$\halmos$\medskip}

\newcommand{\bs}{\begin{split}}
\newcommand{\es}{\end{split}}

\newcommand{\edo}{\end{document}}

\newcommand{\xo}{x^0}

\newcommand{\wo}{w^0}
\newcommand{\IM}{\Sigma _{\mbox{\sc im}}}

\begin{document}

\title{Adaptation and Regulation with Signal Detection Implies Internal Model}

\author{Eduardo D.\ Sontag\thanks{%
E-mail: {\tt sontag@control.rutgers.edu}.
Supported in part by US Air Force Grant
F49620-01-1-0063 and NIH Grant P20 GM64375}\\
Department of Mathematics\\
Rutgers University, New Brunswick, NJ 08903}

\date{}

\maketitle

\begin{abstract}
\noindent
This note provides a simple result showing, under suitable technical
assumptions, that if a system $\Sigma $ adapts to a class of external signals ${\cal U}$,
in the sense of regulation against disturbances or tracking signals in ${\cal U}$,
then $\Sigma $ must necessarily contain a subsystem which is capable of generating
all the signals in ${\cal U}$.
It is not assumed
that regulation is robust, nor is there a prior requirement for the system to
be partitioned into separate plant and controller components.
Instead, one assumes that a ``signal detection'' property holds.
\end{abstract}


\section{Introduction}

\begin{figure}[h,t]
\begin{center}
\setlength{\unitlength}{2000sp}%
\begin{picture}(2475,944)(3526,-3083)
\thicklines
\put(4501,-3061){\framebox(900,900){}}
\thinlines
\put(4051,-2611){\vector( 1, 0){450}}
\put(5401,-2611){\vector( 1, 0){450}}
\put(4876,-2686){$\Sigma $}
\put(6001,-2651){$y(t)\rightarrow 0$}
\put(2600,-2651){$u(\cdot )\in {\cal U}$}
\end{picture}
\caption{Given System, Regulated Output $y(t)$ when Inputs in ${\cal U}$}
\label{fig-sys}
\end{center}
\end{figure}
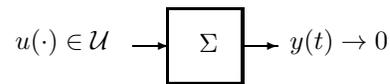

Suppose that one knows that a certain system $\Sigma $ regulates against all those
external input signals $u$ which belong to a predetermined class ${\cal U}$ of
time-functions. 
(Input signals $u$ are often thought of as
disturbances to be rejected or signals to be tracked, depending on the
application.)  In biology, one often uses the term {\em adaptation} for this
property.
It means that a certain quantity $y(t)$ associated to
the system, called its output (also called a regulated variable or an error)
has the property that $y(t)\rightarrow 0$ as $t\rightarrow \infty $ whenever the system is
subject to an input signal from the class ${\cal U}$ (Figure~\ref{fig-sys}).
The {\em internal model principle (IMP)\/} states, roughly, that
the system $\Sigma $  necessarily must contain a subsystem $\IM$ which can itself
generate all disturbances in the class ${\cal U}$.
The terminology arises when thinking of $\IM$ as a ``model'' of
a system which generates the external signals.

For example, if $y(t)\rightarrow 0$ as $t\rightarrow \infty $ whenever the system is subject to any
external constant signal (i.e., the class ${\cal U}$ consists of all constant
functions), then the system $\Sigma $ must contain a subsystem $\IM$
which generates all constant signals (typically an integrator, since
constant signals are generated by the differential equation $\dot u=0$).  
Of course, the choice of $y=0$ as the ``adaptation value'' is merely a matter
of convention; by means of a change of variables, one may always reduce a given
regulation objective ``$y(t)\rightarrow y_0$'' where $y_0$ is some predetermined value,
to the special case $y_0=0$.

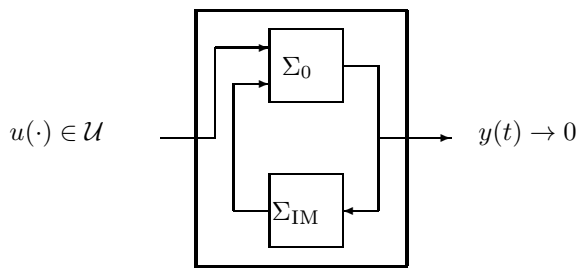
\begin{figure}[h,t]
\begin{center}
\setlength{\unitlength}{3000sp}
\begin{picture}(3150,2144)(2776,-3533)
\thinlines
\put(4201,-2161){\framebox(600,600){}}
\put(4201,-3361){\framebox(600,600){}}
\put(5101,-1861){\line( 0,-1){1200}}
\put(4801,-1861){\line( 1, 0){300}}
\put(5101,-3061){\vector(-1, 0){300}}
\put(5101,-2461){\vector( 1, 0){600}}
\put(4201,-3061){\line(-1, 0){300}}
\put(3901,-2011){\vector( 1, 0){300}}
\put(3301,-2461){\line( 1, 0){450}}
\put(3751,-2461){\line( 0, 1){750}}
\put(3751,-1711){\vector( 1, 0){450}}
\put(3901,-3061){\line( 0, 1){1050}}
\thicklines
\put(3601,-3511){\framebox(1725,2100){}}
\put(5926,-2481){$y(t)\rightarrow 0$}
\put(2050,-2481){$u(\cdot )\in {\cal U}$}
\put(4300,-1936){$\Sigma _0$}
\put(4220,-3136){$\IM$}
\end{picture}
\caption{Decomposition of $\Sigma $ into $\Sigma _0$ and $\IM$, the Latter Driven by
$y(t)$} 
\label{fig-imp}
\end{center}
\end{figure}

In addition, the IMP specifies that, in an appropriate sense, the subsystem
$\IM$ must only have $y$ as its external input, receiving no other direct
information from other parts of the system nor the input signal $u$.
One intuitive interpretation is that $\IM$ generates its ``best guess'' of the
external input $u$ based on how far the output $y$ is from zero.
Pictorially, if we have the situation shown in Figure~\ref{fig-sys}, then
there must be a decomposition of the system $\Sigma $ into two parts, as shown in
Figure~\ref{fig-imp},
where the system $\IM$ (with $y\equiv 0$) is capable of reproducing all the
functions in ${\cal U}$.

The internal model principle originates in the biological cybernetics
literature.  But, as with any ``principle'' in control theory (like dynamic
programming, the maximum principle, etc.) and more generally in mathematics,
the IMP is not a theorem but rather a ``mold'' for many possible theorems,
each of which will hold under appropriate technical assumptions, and whose
conclusions will depend upon the precise meaning of ``class of external
signals'', ``reproducing all functions'', and so on.

The best known instance of an internal model {\em theorem} is due to Francis
and Wonham, who in a series of beautiful and deep papers in the mid 1970s
proved a theorem for linear systems which showed, in essence, that
structurally stable or ``robust'' adaptation forces the existence of 
embedded internal
models.  Partial generalizations of their work to nonlinear
systems were later obtained by Wonham and Hepburn, 
see~\cite{francis-wonham},\cite{wonham77},\cite{wonham-book},\cite{hepburn}-\cite{wonham-hepburn}.
The Francis/Wonham theory applies to systems $\Sigma $
which are already partitioned into a ``plant'' plus a ``controller''.  The
robustness assumption amounts to the requirement that the given controller
should perform appropriately (in the sense that the regulation objective
$y(t)\rightarrow 0$ is achieved) even when the plant 
subsystem -- but most definitely not the controller subsystem -- is arbitrarily
perturbed. 
The conclusion is that the controller is driven by $y$ and incorporates a
model of the external signals.
That some additional condition -- such as structural stability -- must be
imposed is obvious, since the system $\Sigma $ which simply outputs $y=0$ for every
possible input signal $u$ does not contain any subsystem generating the
signals $u$.
We will impose instead a condition which amounts to a {\em signal detection}
property: the output must reflect sudden changes in the input.

Note that this type of objective is very different from what would be typical
in control design: in the latter field, one would ideally not even notice
disturbances (for instance a change in the road grade, in an automobile's
cruise-control system, or a bump in the road, in an active suspension system)
In contrast, in biological applications, signal detection is often an
objective, to be followed by a return to default values.  This subtle
difference in 
desired behaviors, while dealing with what are otherwise similar problems, is
characteristic of many applications of control-theory ideas in biology.

Recent work in molecular biology, cf.\ \cite{taumu}, has suggested that the
IMP could help guide experimentalists and modelers: if certain
characteristics of a system adapt to signals in a given class
(in all the examples so far, constant inputs, such as
for instance
$y(t)=$ the relative ``activity'' of enzymes controlling motors in {\em E.coli}
chemotaxis, with respect to $u(t)=$ concentration of extracellular ligand, but similar considerations may apply to periodic inputs and circadian
 clocks as internal models of day/night periods)
then the IMP could, in principle, help distinguish among mathematical models
which do or do not contain internal models.

With a view toward such biological applications, it is desirable to have
available a theorem which (a) applies to nonlinear systems $\Sigma $, at least
under reasonable technical assumptions, and (b) does not require the system
$\Sigma $ to be split between ``plant'' and ``controller'' subsystems, nor 
(c) requires
structural stability (robustness) in the sense of the Francis/Wonham theory
(which would imply, in the case of the {\em E.coli} motor control network,
that the system should perfectly adapt even if there are arbitrary direct
connections between the external ligands and the motor signals, a matter which
seems difficult to check experimentally), and relies instead upon a signal
detection property.  We present one very elementary
and self-contained such result in this note.
It basically just picks from and ``repackages'' some of
the basic concepts and techniques developed by previous researchers for the
same problems, in
particular: the use of differential geometric techniques and
``output-zeroing'' sets~(\cite{wonham77}, \cite{hepburn}-\cite{wonham-hepburn},
\cite{isidori-byrnes}), dynamical systems notions like omega-limit
sets~(\cite{hepburn},\cite{wonham-hepburn-gnv},\cite{isidori-byrnes}),
and system decompositions motivated by the Center Manifold
Theorem~(\cite{hepburn},\cite{wonham-hepburn-jmaa},\cite{wonham-hepburn-gnv},\cite{isidori-byrnes}). 
Isidori's
excellent textbooks~\cite{isidori1},\cite{isidori2} should be consulted for a
far deeper discussion of many of the issues raised here.

Precise mathematical definitions are provided in Section~\ref{sec-defs}.
On the other hand, since the linear version of the result is very easy to
explain, we sketch that case first.
(The discussion assumes some familiarity with frequency domain techniques, and
may be skipped without loss of continuity.)

\subsection{Linear Case}
\label{sec-lin-transfer-functions}

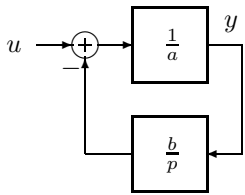
\begin{figure}[h,t]
\begin{center}
\setlength{\unitlength}{2000sp}%
\begin{picture}(2937,2294)(2926,-4433)
\thinlines
\put(3901,-2611){\circle{336}}
\thicklines
\put(4501,-3061){\framebox(900,900){}}
\put(4501,-4411){\framebox(900,900){}}
\thinlines
\put(5401,-2611){\line( 1, 0){450}}
\put(5851,-2611){\line( 0,-1){1350}}
\put(5851,-3961){\vector(-1, 0){450}}
\put(4051,-2611){\vector( 1, 0){450}}
\put(4501,-3961){\line(-1, 0){600}}
\put(3901,-3961){\vector( 0, 1){1200}}
\put(3301,-2611){\vector( 1, 0){450}}
\put(3901,-2536){\line( 0,-1){150}}
\put(3826,-2611){\line( 1, 0){150}}
\put(4876,-2686){$\frac{1}{a}$}
\put(5626,-2386){$y$}
\put(3600,-2986){$-$}
\put(2926,-2700){$u$}
\put(4876,-4036){$\frac{b}{p}$}
\end{picture}
\caption{System Equivalent to $\Sigma $: Closed-loop Zeros are Feedback Poles}
\label{fig-zero-dyn}
\end{center}
\end{figure}

Let us denote by $S$ 
the {\em transfer function} of the system $\Sigma $:
if $y$ is the output produced when $\Sigma $ starts at the zero initial state
and is fed input $u$, then the relation $\hat y(s)=S(s)\hat u(s)$ holds
between the Laplace transforms $\hat y(s)$, $\hat u(s)$ of the output and
input.
One expresses $S(s)=\frac{p(s)}{q(s)}$ as the quotient of two relatively prime
polynomials, with the degree of $p$ less than the degree of $q$.
(An equivalent discussion using differential operators instead of Laplace
transforms is also possible, see e.g.\ Section 6.7 in~\cite{mct}.)
The first observation, a well-known fact in systems theory, is that the zeroes
of $p$ can be viewed, alternatively, as poles of a feedback subsystem.
To see this, we assume that $p$ is not identically zero, and divide the
polynomial $q$ by $p$, obtaining $q=a p + b$, where $b$ is some polynomial of
degree less than $p$.
Now, as the algebraic equality $y=\frac{p}{q}u$ is equivalent to
$y = \frac{1}{a}(u-\frac{b}{p}y)$, we conclude that the system $\Sigma $ can be
decomposed as in Figure~\ref{fig-zero-dyn}.
For example, if $s=0$ is a zero of $S$ (that is, $0$ is a root of $p$), which
amounts to the property that
constant signals get differentiated by $\Sigma $ (the ``DC gain'' of $\Sigma $ is zero),
then the factor $1/s$ appears in the feedback box $b/p$, and can be
interpreted as an  integrator of the output $y$.

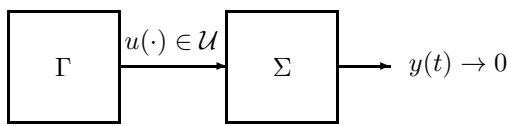
\begin{figure}[h,t]
\begin{center}
\setlength{\unitlength}{3000sp}
\begin{picture}(3322,944)(2679,-3083)
\thicklines
\put(4501,-3061){\framebox(900,900){}}
\put(2701,-3061){\framebox(900,900){}}
\thinlines
\put(5401,-2611){\vector( 1, 0){450}}
\put(4876,-2686){$\Sigma $}
\put(6001,-2651){$y(t)\rightarrow 0$}
\put(3650,-2461){$u(\cdot )\in {\cal U}$}
\put(3601,-2611){\vector( 1, 0){900}}
\put(3076,-2686){$\Gamma $}
\end{picture}
\caption{Exosystem and System in Cascade}
\label{fig-sys-and-generator}
\end{center}
\end{figure}

We will show that, in general, the subsystem with transfer function
$\frac{b}{p}$ models all inputs which $\Sigma $ adapts to.
Let us suppose that the class ${\cal U}$ of inputs can be described as the set of
all possible solutions of a fixed linear differential equation
\[
u^{(\ell)}(t) + b_1 u^{(\ell-1)}(t) + \ldots  + b_{\ell-1} u'(t) + b_\ell u(t)
\;=\; 0 
\]
for some integer $\ell$, and 
which has no stable modes.  (Stable modes, giving components of $u$ which
converge to zero, are less interesting, since they do not represent persistent
disturbances.) 
We view these signals $u$ as the outputs of an ``exosystem'' $\Gamma $ which is
obtained by rewriting the differential equation as a system of $\ell$ first
order equations.
Figure~\ref{fig-sys-and-generator} shows a cascade consisting of the original
system $\Sigma $ and the exosystem $\Gamma $ which generates the inputs in ${\cal U}$.
(If, for example, ${\cal U}=$ constant inputs, then one would let $\Gamma $ be the system
with equation $\dot w=0$ and output $u=w$, and for each initial condition
$w(0)=u^0$, one obtains a different constant output $u(t)\equiv u^0$.)
The regulation objective is now simply that $y(t)\rightarrow 0$ for all possible initial
conditions of the composite system, i.e.\ for all initial conditions of the
original system $\Sigma $ and all initial conditions of the exosystem $\Gamma $, the
latter corresponding to all possible inputs in ${\cal U}$ being fed to $\Sigma $.

\begin{figure}[h,t]
\begin{center}
\setlength{\unitlength}{3000sp}
\begin{picture}(4350,944)(1651,-3083)
\thicklines
\put(4501,-3061){\framebox(900,900){}}
\put(2701,-3061){\framebox(900,900){}}
\thinlines
\put(5401,-2611){\vector( 1, 0){450}}
\put(4876,-2686){$\Sigma $}
\put(6001,-2651){$y(t)\rightarrow 0$}
\put(3601,-2611){\vector( 1, 0){900}}
\put(3076,-2686){$\Gamma $}
\put(2251,-2611){\vector( 1, 0){450}}
\put(1200,-2651){$v(t)\rightarrow 0$}
\end{picture}
\caption{Exosystem and System, Forced by Stable Inputs}
\label{fig-sys-and-generator-inputs}
\end{center}
\end{figure}
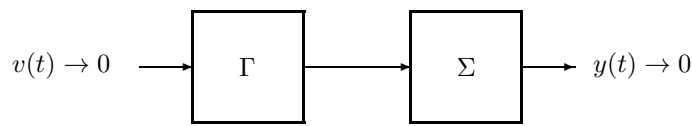

\begin{figure}[h,t]
\begin{center}
\setlength{\unitlength}{2000sp}%
\begin{picture}(3354,1084)(4009,-4433)
\thicklines
\put(4501,-4411){\framebox(900,900){}}
\put(6001,-4411){\framebox(900,900){}}
\thinlines
\put(7351,-3361){\line( 0,-1){600}}
\put(7351,-3961){\vector(-1, 0){450}}
\put(4501,-3961){\line(-1, 0){450}}
\put(4051,-3961){\vector( 0, 1){600}}
\put(6001,-3961){\vector(-1, 0){600}}
\put(4750,-4036){$\frac{b_1}{p_0}$}
\put(6300,-4036){$\frac{b_2}{\pi }$}
\end{picture}
\caption{Decomposition of $\frac{b}{p}$}
\label{fig-zero-dyn-decomp}
\end{center}
\end{figure}
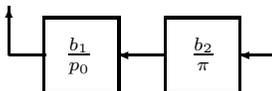

Next, we reformulate this regulation property by adding an external input 
$v(\cdot )$ to the exosystem, and requiring now that $y(t)\rightarrow 0$ for all possible
stable inputs ($v(t)\rightarrow 0$ as $t\rightarrow \infty $) but only when starting from the zero
initial state.
(Such replacements of initial states by stable forcing inputs -- assuming
natural controllability/observability 
conditions -- are elementary exercises in linear systems theory, see e.g.\ the
proof of Theorem 33 in~\cite{mct}.) 
In other words, we have now the situation illustrated by
Figure~\ref{fig-sys-and-generator-inputs}.
We denote by $G$ the transfer function of the exosystem $\Gamma $:
$G=\frac{1}{\pi }$, where
\[
\pi (s)\;=\;
s^{\ell} + b_1 s^{\ell-1} + \ldots  + b_{\ell-1} s + b_\ell \,.
\]
To see that the subsystem with transfer function ${b}/{p}$ includes an
internal model of $\Gamma $, we argue as follows.
The regulation property for the cascade in
Figure~\ref{fig-sys-and-generator-inputs} means that the product rational
function $GS$ is stable (all poles have negative real parts), while the
assumption that $G$ had no stable modes means that all the poles of $G$
(i.e, the roots of the polynomial $\pi $) have nonnegative real parts.  
Therefore, these poles must be canceled in the product $GS$; in other words,
$S$ {\em must have among its zeroes all the poles of\/} $G$, so
that we can write $p=\pi  p_0$ for some polynomial $p_0$.
Thus ${b}/{p} = {b}/({\pi  p_0})$.
One may now factor $b=b_1 b_2$ in such a way that the degree of $b_2$ is less
than the degree of $\pi $, so that ${b}/{p} = (b_1/p_0)(b_2/\pi )$ and now the
system with transfer function ${b}/{p}$ can be written itself in the cascade
form in Figure~\ref{fig-zero-dyn-decomp}.
The subsystem with transfer function $b_2/\pi $ generates all the inputs
in ${\cal U}$, since one may write a set of differential equations for it which is
exactly the same as for the exosystem $\Gamma $, changing only the output
mapping (``controller form'' realization).

Since the tools of transfer functions are not available for nonlinear systems,
a different approach is required in general.

\section{Definitions and Statement of Result}
\label{sec-defs}

We consider single-input single-output systems $S$, affine in inputs:
\be{sys}
\dot x(t)=f(x(t)) + u(t) g(x(t))\,,\;y(t)=h(x(t))
\ee
(dot indicates derivative with respect to time, and the arguments $t$ will be
omitted from now on; see~\cite{mct} for general definitions and properties of
systems with inputs).
Here $x(t)$, $u(t)$, and $y(t)$ represent the state, input, and output at time
$t$, $f$ and $g$ are smooth vector fields on $\R^n$ ($n$ is the dimension of
the system), $h$ is a scalar smooth function $\R^n\rightarrow \R$, and $f(0)=h(0)=0$.
(Several assumptions on $f$ and $g$ will be made later.)
A special case is that of {\em linear\/} systems
\be{sys-linear}
\dot x=Ax+ub\,,\;y=cx
\ee
where $A$ is an $n\times n$ matrix, $b$ is a column $n$-vector, and $c$ is
a row $n$-vector.

Suppose given a class ${\cal U}$ of functions $[0,\infty )\rightarrow \R$ (such as for example the
set of all constant functions).
We say that $\Sigma $ {\em adapts} to inputs in ${\cal U}$ (a more appropriate technical
control-theoretic term would be ``asymptotically rejects disturbances in
${\cal U}$'') if the following property holds: for each $u\in {\cal U}$ and each initial
state $\xo\in \R^n$, the solution of~(\ref{sys}) with initial condition
$x(0)=\xo$ 
exists for all $t\geq 0$ and is bounded, and the corresponding output
$y(t)=h(x(t))$ converges to zero as $t\rightarrow \infty $.

We will say that $S$ {\em contains an output-driven internal model of ${\cal U}$}
if there is a change of coordinates which brings the equations~(\ref{sys})
into the following block form:
\begin{equation}
\label{sys-with-im}
\begin{array}{rcl}
\dot z_1 &=& f_1(z_1,z_2) + u g_1(z_1,z_2)  \\
\dot z_2 &=& f_2(y,z_2) \\
    y &=& \kappa (z_1)
\end{array}
\end{equation}
(the subsystems with variables $z_1$ and $z_2$ correspond respectively to
$\Sigma _0$ and $\IM$ in Figure~\ref{fig-imp}),
and in addition the subsystem with state
variables $z_2$ is capable of generating all functions in ${\cal U}$, meaning the
following property: there is some scalar function $\varphi(z_2)$ so that, for each
possible $u\in {\cal U}$, there is 
some solution of
\be{z2sys}
\dot z_2 = f_2(0,z_2)
\ee
which
satisfies $\varphi(z_2(t))\equiv u(t)$.

The precise meaning of ``change of coordinates'' is as follows.
There must exist an integer $r\leq n$, differentiable manifolds $Z_1$ and $Z_2$
of dimensions $r$ and $n-r$ respectively, a smooth function $\kappa :Z_1\rightarrow \R$,
vector fields $F$ and $G$ on $Z_1\times Z_2$ which take the partitioned form
\[
F = \begin{pmatrix}f_1(z_1,z_2)\cr f_2(\kappa (z_1),z_2)\end{pmatrix}
\,,\quad
G = \begin{pmatrix}g_1(z_1,z_2)\cr 0\end{pmatrix}
\]
and a diffeomorphism $\Phi :\R^n\rightarrow Z_1\times Z_2$, such that
\[
\Phi _*(x)f(x) = F(\Phi (x))\,,\quad
\Phi _*(x)g(x) = G(\Phi (x))\,,\quad
\kappa (\Phi _1(x))=h(x)
\]
for all $x\in \R^n$, where $\Phi _1$ is the $Z_1$-component of $\Phi $ and star
indicates Jacobian.

Our result will hold under additional conditions on the vector fields defining
the system.  The first condition is the fundamental one from an intuitive
point of view, namely that the system is able to detect changes in the input
signal:
\begin{center}
\framebox{{\bf Assumption 1:}
a uniform relative degree exists.}
\end{center}
This means that there exists some positive integer $r$ such that
\[
L_gL_f^kh\equiv 0\quad\forall\, k<r-1
\]
and
\[
L_gL_f^{r-1}h(x)\not= 0\quad\forall\,x\in \R^n
\]
where, as usual, $L_Xh$ indicates the directional derivative (``Lie
derivative'') of a function $h$ along the direction of the vector field
$X$, that is $(L_Xh)(x) = \nabla h(x)\cdot X(x)$.
The integer $r$, if it exists, is called the {\em relative degree} of $\Sigma $.
It is possible to prove (see~\cite{isidori1}) that when $r$ exists, necessarily
$r\leq n$.

For a linear system~(\ref{sys-linear}), existence of a relative degree amounts
to simply asking that $cA^ib$ is nonzero for some $i$, or equivalently that
the transfer function $c(sI-A)^{-1}b$ is not identically zero.
For general systems~(\ref{sys}), the assumption is equivalent to the statement
that the output derivatives $y^{(k)}(t)$ must be independent of the value of
the input at time $t$, for all $k<r$, but that
$y^{(r)}(t)=b(x(t))+a(x(t))u(t)$ for some function $a(x)$ which is everywhere
nonzero (so that the system can be ``inverted'' to obtain the instantaneous
value $u(t)$ from instantaneous derivatives).
See also~\cite{lms} for a discussion of the characterization of $r$ in terms
of smoothness of outputs when inputs are discontinuous (change detection).

The next two conditions are of a technical nature.
They are automatically satisfied
for linear systems.  For nonlinear systems, we need such conditions in order
to guarantee the existence of a change of variables exhibiting the system
$\IM$.  (See Remark~\ref{weak-rem} for ways of weakening these assumptions.) 
We are guided by conditions which appear in Isidori's book~\cite{isidori1}.

Assuming that the degree is $r$, we introduce the following vector fields:
\[
\widetilde g(x) \;=\; \frac{1}{L_gL_f^{r-1}h(x)}g(x)\,,\quad
\widetilde f(x) \;=\; f(x) - \Big(L_f^rh(x)\Big) \widetilde g(x)\,,\quad
\tau _i:={\rm ad}^{i-1}_{\widetilde f} \widetilde g ,\;i=1,\ldots r\,,
\]
where ${\rm ad}_X$ is the
operator ${\rm ad}_X Y = [X,Y]=$ Lie bracket of the vector fields $X$ and $Y$.
Recall that a vector field $X$ is said to be {\em complete} if the solution
of the initial value problem $\dot x=X(x)$, $x(0)=\xo$ is defined for all $t\in \R$,
for any initial state $\xo$, and that two vector fields $X$ and $Y$ are said to
{\em commute} if $[X,Y]=0$.
The assumptions are:
\begin{center}
\framebox{{\bf Assumption 2:}
$\tau _i$ is complete, for $i=1,\ldots ,r$.}
\end{center}
\begin{center}
\framebox{{\bf Assumption 3:}
the vector fields $\tau _i$ commute with each other.}
\end{center}

Finally, we must define the allowed classes of inputs ${\cal U}$.
As usual in control theory
(see also the discussion in Section~\ref{sec-lin-transfer-functions}), we will
assume that inputs are generated by {\em exosystems}.  That is, there is given
a system $\Gamma $:
\be{exosys}
\dot w = Q(w)\,,\quad u = \theta (w)
\ee
(let us say evolving on some differentiable manifold, $Q$ a smooth vector
field, and $\theta $ a real-valued smooth function, although far less than
smoothness is needed) such that the input class ${\cal U}$ consists exactly of those
inputs $u(t)=\theta (w(t))$, $t\geq 0$, for all possible solutions of $\dot w=Q(w)$.
For example, if we are interested in constant signals, we pick
$\dot w=0$, $u=w$
and if we are 
interested in sinusoidals with frequency $\omega $ then we use
$\dot x_1=x_2$, $\dot x_2=-\omega ^2x_1$, $u=x_1$.
It is by now standard in nonlinear studies of necessary conditions for
regulation to impose conditions on omega limits sets for trajectories of the
exosystem, see~\cite{hepburn},\cite{wonham-hepburn-gnv}; we will follow the
approach in~\cite{isidori-byrnes}-\cite{isidori1} and
assume that the exosystem 
is Poisson-stable: for every state $w^0$, the solution
$w(\cdot )$ of $\dot w=Q(w)$, $w(0)=w^0$ is defined for all $t>0$ and it satisfies
that $w^0$ is in the omega-limit set of $w$, that is, there is a sequence of
times $t_i\rightarrow \infty $ such that the sequence $w(t_i)$ converges to $w^0$ as
$t\rightarrow \infty $.
This means that the exosystem is almost-periodic in the sense that
trajectories keep returning to neighborhoods of the initial state.

This theorem is proved in Section~\ref{sec-proof}:

\bt{main-theo}
If Assumptions 1-3 hold and the system $\Sigma $ adapts to inputs in a class
${\cal U}$ generated by a Poisson-stable exosystem, then
$S$ contains an output-driven internal model of ${\cal U}$.
\ets

\subsection{An Example}

As an example, consider the model for {\em E.coli}
chemotaxis adaptation to constant inputs
given in~\cite{iglesias-levchenko}, Section 2.2.
Letting $x_1=R$ and $x_2=RL$ be the concentrations of unbound and bound
receptors respectively, and taking the external ligand concentration $u=L$ as
input, we have the following equations:
\begin{equation}
\label{iglesias-chemotaxis}
\begin{array}{rcl}
\dot x_1 &=& a_1 - a_2 x_1 + a_3 x_2 - a_4 x_1 u \\
\dot x_2 &=& a_5 -  a_6 x_2 + a_4  x_1 u
\end{array}
\end{equation}
for suitable positive constants $a_1,\ldots ,a_6$.
In terms of vector fields,
\[
f\;=\;
\begin{pmatrix} 
a_1 - a_2 x_1 + a_3 x_2 \cr a_5 -  a_6 x_2 
\end{pmatrix}
\,,\quad
g\;=\;
\begin{pmatrix} 
- a_4 x_1 \cr a_4  x_1
\end{pmatrix}
\]
and, still as in~\cite{iglesias-levchenko}, we take as output $y$ the
difference between the total concentration of active receptors and a
steady state level of this activity.
In terms of the notations used here, and up to multiplication by a suitable
constant, this amounts to the following choice:
\[
h(x) \;=\; A_0 - A \;=\; [a_1+a_5] - [a_2 x_1+(a_6-a_3)x_2] \,.
\]
We note that  $L_gh = D x_1$, where $D= a_2 a_4+(a_3-a_6)a_4$.
Except in the accidental case when this constant $D$ vanishes
(in terms of the notations in~\cite{iglesias-levchenko},
$D=k_{-1}I_T k_r(\alpha _1-\alpha _2)$, so $D$ can only vanish if $\alpha _1=\alpha _2$),
we have that $L_gh(x)\not= 0$ for all $x$ ($x_1>0$, as it represents a
concentration), and so it follows that $\Sigma $ has well-defined relative degree
$r=1$.
Moreover, $\tau _1=\widetilde g$ is a constant vector, so Assumptions 2 and
3 hold as well.

A minor technicality concerns the assumptions that our systems~(\ref{sys})
evolve in all of Euclidean state space (not just $x_i>0$)
and that $f(0)=h(0)=0$.
However, this is just a matter of picking the right coordinates.
Notice that $f$ vanishes at $x^0=(x_1^0,x_2^0)$, where
$x_1^0=(a_1a_6+a_3a_5)/(a_2a_6)$ and
$x_2^0=a_5/a_6$,
and $h$ vanishes at $x^0$ too.
In order to fit into the general theory, one simply changes variables,
mapping the positive orthant into all of $\R^2$ and $x^0$ into the origin
by means of $x_i'=\ln x_i - \ln x_i^0$.
(Of course, there is no need to actually perform the coordinate change, since
conditions expressed in terms of Lie derivatives are covariant.)

Finally, letting $B:=x_1+x_2$ (as done in~\cite{iglesias-levchenko}), one
obtains a system of equations in terms of the new variables $A$ and $B$, for
which $\dot B = y$.  This last equation represents an integrator (internal model
of a system which produces constant inputs) driven by the output $y$.
(Of course, there is no point in applying the theorem, since once that the
model is given we can find the internal model explicitly.)

\section{Proof of Theorem~\protect{\ref{main-theo}}}
\label{sec-proof}

Suppose that the system $\Sigma $ adapts to inputs in ${\cal U}$, which are produced by a
Poisson stable exosystem $\Gamma $.
We consider the interconnected system consisting of the cascade of $\Gamma $ and
$\Sigma $, as shown in Figure~\ref{fig-sys-and-generator}, namely:
\begin{equation}
\label{composite-sys-general}
\begin{array}{rcl}
\dot w &=& Q(w)\\
\dot x &=& f(x) \,+\, \theta (w) g(x)
\end{array}
\end{equation}
and let ${\cal Z}$ denote the set consisting of those states $x$ of $\Sigma $ for which
$h(x)=0$ (the ``output-zeroing'' subset).

\bl{main-lemma}
For each $\wo$ there is some solution $\sigma =(w(\cdot ),x(\cdot ))$ of the composite
system~(\ref{composite-sys-general}) such that $w(0)=\wo$ and $x(t)\in {\cal Z}$ for
all $t\geq 0$.
\els

\bpr
We start by picking an arbitrary solution $\sigma _0=(w(\cdot ),x(\cdot ))$ of the
composite system~(\ref{composite-sys-general}) such that $w(0)=\wo$,
and let $\Omega  = \Omega ^+[\sigma _0]$ be the omega-limit set of this trajectory.
We claim that, for each point $(\omega ,\xi )\in \Omega $ (we partition coordinates into
those for $\Gamma $ and $\Sigma $) it must be the case that $\xi \in {\cal Z}$.
Indeed, by definition of $\Omega $ there is some sequence of times $t_i\rightarrow \infty $ such
that $x(t_i)\rightarrow \xi $.  Since $h(x(t_i))\rightarrow 0$ because of the adaptation property and
$h$ is continuous, it follows that $h(\xi )=0$, as claimed.
Next, we claim that there is some $\xo$ such that $(\wo,\xo)\in \Omega $.
To see this, we first pick a sequence of times $t_i\rightarrow \infty $ such
that $w(t_i)\rightarrow \wo$ (Poisson stability is used here); as $\{x(t_i)\}$ is
bounded, we may pick a subsequence $t_{i_j}$ of the $t_i$ so that
$x(t_{i_j})\rightarrow \xo$ for some $\xo$, and this proves that $(\wo,\xo)\in \Omega $, as
wanted.

Finally, we let $\sigma $ be the solution $\sigma =(w(\cdot ),x(\cdot ))$ of the composite
system~(\ref{composite-sys-general}) for which $w(0)=\wo$ and $x(0)=\xo$,
where $\xo$ is so that $(\wo,\xo)\in \Omega $.
Omega-limit sets are invariant, so $\sigma (t)\in \Omega $ for all $t\geq 0$, and
we already proved that this last property implies that $x(t)\in {\cal Z}$.
\epr

Proposition 9.1.1 in~\cite{isidori1} shows that there is a global
diffeomorphism $\Phi $ so that, in the new coordinates, the system $\Sigma $ takes the
form shown in Display~(\ref{sys-with-im}).
Moreover, the subsystem described by $z_1$ evolves in $\R^r$ and, using
coordinates $z_1=(\zeta _1,\ldots ,\zeta _r)$, the equations for $z_1$ can be written as
follows:
\begin{equation}
\label{derivs-of-y-system}
\begin{array}{rcl}
\dot \zeta _1    &=& \zeta _2  \\
          &\vdots&  \\
\dot \zeta _{r-1} &=& \zeta _r  \\
\dot \zeta _r     &=& b(z_1,z_2) + a(z_1,z_2) u
\end{array}
\end{equation}
where the output is $y=\kappa (z_1)=\zeta _1$ and $a,b$ are smooth functions with
$a(z) = L_gL_f^{r-1}h(\Phi ^{-1}(z))\not= 0$ for all $z$.
We let
\[
\varphi(z_2)\,:=\; -\frac{b(0,z_2)}{a(0,z_2)}
\]
and show that for each possible $u\in {\cal U}$
there is  some solution of~(\ref{z2sys})
which satisfies $\varphi(z_2(t))\equiv u(t)$.

We pick $\wo$ such that $u(t)=\theta (w(t))$ and $w(0)=\wo$, and
view the interconnection~(\ref{composite-sys-general}) of $\Gamma $ and $\Sigma $
in terms of the coordinate change given by $\Phi $ on $\Sigma $:
\beqn
\dot w &=& Q(w)\\
\dot z_1 &=& f_1(z_1,z_2) \,+\, \theta (w) g_1(z_1,z_2)  \\
\dot z_2 &=& f_2(y,z_2) \,.
\eeqn
Lemma ~\ref{main-lemma} gives us the existence of
a solution $\sigma =(w(\cdot ),z_1(\cdot ),z_2(\cdot ))$ such that $\theta (w(t))=u(t)$ and
$\zeta _1(t)\equiv 0$.
Because of the form~(\ref{derivs-of-y-system}) of the $z_1$-subsystem, this
implies that $z_1(t)\equiv 0$ and that $\dot \zeta _r(t)\equiv 0$.
Thus, along the solution $\sigma $ one has $b(0,z_2(t)) + a(0,z_2(t)) u(t)\equiv 0$,
and this is precisely what we wished to prove.
\qed

\br{weak-rem}
Assumptions 2 and 3 are automatically satisfied for linear systems, since
the vector fields $\tau _i$ are all constant, so that they
are indeed complete and pairwise commutative.
For general nonlinear systems, these assumptions, especially 3, are quite
strong.
Weaker conditions may be given, if one is merely interested in a
local result, or if one is willing to accept a subsystem $\IM$ which is driven
by not just $y$ but also several derivatives of $y$.
Indeed, assuming merely a well-defined relative degree around a given point
$x^0$, we obtain a decomposition as in~(\ref{sys-with-im})
(see Section 4.3 in~\cite{isidori1})
except that the change of coordinates is now only valid in a neighborhood of
$x^0$, and $f_2$ now depends on $(z_1,z_2)$ (as opposed to $(y,z_2)$.
Note that, from the form~(\ref{derivs-of-y-system}), $z_1$ is the vector
consisting of the derivatives $y, y',\ldots  y^{(r)}$. 
If condition 2 holds, but 3 does not, then a global decomposition is possible,
but $f_2$ still depends on derivatives of $y$ (cf.\ \cite{isidori1},
Proposition 9.1.1).
\er

\subsubsection*{A Remark on Subsystems}
\label{sec-remark-subsystem}

We expressed our theorem in terms of the existence of solutions which reproduce
all inputs.  Under additional and stronger hypotheses, one could also
obtain an actual embedding of the exosystem in the internal model $\IM$.
A full nonlinear version would involve abstract quotients of systems
under suitable equivalence relations, and may follow along the lines of the
work in~\cite{wonham-hepburn-mct} (based on~\cite{hermann-krener}).
However, the necessary steps are easy to understand and prove in the case of
{\em linear\/} systems.
We start by showing the following elementary fact from linear
systems theory

\bl{lin-sys-lemma}
Suppose given an observable linear system $\dot w=Qw$, $y=\theta  w$ and another
linear system $\dot z_2=F z_2+Gy$, $u=\varphi z$, and assume that for each $w^0$ there
is some $z^0$ such that $\varphi e^{tF}z^0=\theta  e^{tQ}w^0$ for all $t\geq 0$.
Then, the matrix $F$ is similar to a matrix with this block structure:
\be{linear-block-structure}
\begin{pmatrix} 
Q&0&0\cr D&E&0\cr F&G&H
\end{pmatrix} \,.
\ee
\els

\bpr
We first assume that the pair $(F,\varphi)$ is observable, and claim that for each
$w^0$ there is a {\em unique} $z^0$ such that $\varphi e^{tF}z^0\equiv \theta  e^{tQ}w^0$.
This is because $\varphi e^{tF}z^0\equiv \varphi e^{tF}z^1$ implies $z^0=z^1$
(observability).
So we can define a map $T:w^0\mapsto z^0$.
This map is one-to-one, by observability of the pair $(Q,\theta )$.
It is also linear, since 
$\theta  e^{tQ}(\alpha  w^0 + w^1) = 
\alpha  \theta  e^{tQ}w^0 + \theta  e^{tQ} w^1 = 
\alpha  \varphi e^{tF}Tw^0 + \varphi e^{tF} Tw^1 = 
\varphi e^{tF}(\alpha  Tw^0 +Tw^1)$
means that $\alpha  w^0 + w^1 \mapsto  \alpha  Tw^0 +Tw^1$.
It also satisfies $FT=TQ$, since taking derivatives in
$\varphi e^{tF} T w^0\equiv \theta  e^{tQ}w^0$
gives
$\varphi e^{tF}F T w^0\equiv \theta  e^{tQ}Qw^0$
which means that $Qw^0 \mapsto F T w^0$.
Thus, on some invariant subspace (the range of $T$), $F$ can be written as
$Q$, which means that we can write $F$ up to similarity in the form
$\begin{pmatrix}
Q & *\cr 0 & * 
\end{pmatrix}$.
Since $F$ is similar to its transpose, and $Q$ is similar to its transpose,
$F$ is also similar to a matrix in the form
$\begin{pmatrix}Q & 0 \cr * & * \end{pmatrix}$.
An observability decomposition (\cite{mct}, Chapter 6) then reduces to the
observable case.
\epr

Without loss of generality, one may assume that linear exosystems are
observable (there always exists an observable equivalent).
We now apply Lemma~\ref{lin-sys-lemma} to the exosystem and the internal model
$\IM$, assumed linear.
There results a change of variables for $\IM$ so that, in the new variables,
a subset $\zeta $ of the variables $z_2$ of $\IM$, corresponding to the first block
in~(\ref{linear-block-structure}), evolves according to an equation of
the form $\dot \zeta  = Q\zeta  + b y$, for a suitable vector $b$.
This provides the desired embedding of the exosystem in the internal model.

\edo